\documentclass[final]{colt2021}

\title[Quantifying Variational Approximation for Log-Partition Function]{Quantifying Variational Approximation for Log-Partition Function}

\usepackage{times}
\usepackage{amsfonts}
\usepackage{xcolor}
\usepackage{graphicx}
\newcommand{\trw}{\text{\small TRW}}
\newcommand{\maxcut}{\text{\small MAXCUT}}
\newcommand{\maxcsp}{\text{\small MAXCSP}}
\newcommand{\trwp}{\text{\small TRW}^\prime}
\newcommand{\alg}{\text{ALG}}
\newcommand{\rhos}{\rho^\star}
\newcommand{\brhos}{\brho^\star}
\newcommand{\bzero}{{\mathbf 0}}
\newcommand{\bs}{{\mathbf s}}
\newcommand{\bw}{{\mathbf w}}
\newcommand{\bws}{\bw^\star}
\newcommand{\ws}{w^\star}
\newcommand{\Prt}{{\mathsf {Part}}}
\newcommand{\Fs}{F^\star}

\newcommand{\Hs}{{\mathsf H} }

\newcommand{\hu}{\hat{u}}

\newcommand{\bu}{{\mathbf u}}
\newcommand{\ubf}{{\mathbf u}}
\newcommand{\hbu}{\hat{\bu}}

\newcommand{\primal}{\textbf{Primal}}
\newcommand{\dual}{\textbf{Dual}}

\newcommand{\Ptree}{{\sf P}^{\text{tree}}}
\newcommand{\bv}{{\mathbf v}}

\newcommand{\Ccal}{\mathcal{C}}

\newcommand{\Ical}{\mathcal{I}}

\newcommand{\Ncal}{\mathcal{N}}
\newcommand{\Pcal}{\mathcal{P}}

\newcommand{\Tcal}{\mathcal{T}}
\newcommand{\Ucal}{{\mathbf u}}

\newcommand{\Wcal}{\mathcal{W}}
\newcommand{\Xcal}{\mathcal{X}}
\newcommand{\Ycal}{\mathcal{Y}}

\newcommand{\brho}{\boldsymbol{\rho}}

\newcommand{\Ebb}{\mathbb{E}}

\newcommand{\Rbb}{\mathbb{R}}
\newcommand{\Rbbp}{\Rbb_+}

\newcommand{\bX}{{\mathbf X}}
\newcommand{\bx}{{\boldsymbol x}}

\newcommand{\btheta}{\boldsymbol{\theta}}

\newcommand{\Pb}{\mathbb{P}}

\newcommand{\hPhi}{\widehat{\Phi}}

\usepackage{bbm}
\newcommand{\one}{\mathbbm{1}}
\newcommand{\1}{\mathbbm{1}}
\newcommand{\aprx}{\alpha}

\newcommand{\x}{\mathsf{x}}
\newcommand{\T}{\mathsf{T}}

\DeclareMathOperator*{\argmax}{arg\,max}
\DeclareMathOperator*{\argmin}{arg\,min}
\newcommand{\smiddle}[1]{\;\middle#1\;}

\definecolor{dark_red}{rgb}{0.2,0,0}

\makeatletter
\let\Ginclude@graphics\@org@Ginclude@graphics
\makeatother

\coltauthor{%
 \Name{Romain Cosson} \Email{cosson@mit.edu}\\
 \addr Massachusetts Institute of Technology
 \AND
 \Name{Devavrat Shah} \Email{devavrat@mit.edu}\\
 \addr Massachusetts Institute of Technology
}

\begin{document}
\maketitle
\begin{abstract}
Variational methods, such as mean-field (MF) and tree-reweighted (TRW), provide computationally efficient approximations of the log-partition function for generic graphical models but their approximation ratio is generally not quantified.  
As the primary contribution of this work, we provide an approach to quantify their approximation ratio for any discrete pairwise graphical model with non-negative potentials through a property of the underlying graph structure $G$. 
Specifically, we argue that (a variant of) TRW produces an estimate within factor $1/\sqrt{\kappa(G)}$ where $\kappa(G) \in (0,1]$
captures how far $G$ is from tree structure. 
As a consequence, the approximation ratio is $1$ for trees, $\sqrt{(d+1)/2}$ for graphs with maximum average degree $d$ and $1+1/(2\beta)+o_{\beta\to \infty}(1/\beta)$ for graphs with girth at least $\beta \log N$.
The quantity $\kappa(G)$ is the solution of a min-max problem associated with the spanning tree polytope of $G$ that can be evaluated in polynomial time for any graph. 
We provide a near linear-time variant that achieves an approximation ratio depending on the minimal (across edges) effective resistance of the graph. We connect our results to the graph partition approximation method and thus provide a unified perspective.
\end{abstract}

\begin{keywords}%
  variational inference, log-partition function, spanning tree polytope, minimum effective resistance, balanced covering of graph, min-max spanning tree, local inference
\end{keywords}

\section{Introduction}

\medskip
\noindent{\bf The Setup.} 
We consider a collection of $N$ discrete valued random variables on a discrete alphabet, $\bX = (X_1,\dots, X_N)$, 
whose joint distribution is modeled as a pair-wise graphical model. Let $G = (V, E)$ represent 
the associated graph with vertices $V = \{1,\dots, N\}$ representing $N$ variables 
and $E \subset V \times V$ representing edges. Let each variable take value in a 
discrete alphabet $\Xcal$. For $e \in E$,  let $\phi_e: \Xcal \times \Xcal \to \Rbbp$ 
denote the edge potential that we assume takes only non-negative values and let $\theta_e \in \Rbbp$ denote the associated weight. 
This leads to a joint distribution with probability mass function
\begin{align}
    \Pb(\bX = \bx; \btheta) & \propto \exp\Big(\sum_{e \in E} \theta_e \phi_e(x_e)\Big) ~
                   =~ \frac{1}{Z(\btheta)} \exp\Big(\sum_{e \in E} \theta_e \phi_e(x_e)\Big), \label{eq:pairwise}
\end{align} 
where $\bx=(x_1,\dots, x_N) \in \Xcal^N$, $x_e$ is short hand for $(x_s, x_t)$ if $e = (s, t) \in E$, 
$\btheta = (\theta_e: e \in E) \in \Rbbp^{|E|}$ and the normalizing constant or partition function $Z(\btheta)$ is
defined as 
\begin{align}
    Z(\btheta) & = \sum_{\bx \in \Xcal^N} \exp\Big(\sum_{e \in E} \theta_e \phi_e(x_e)\Big). \label{eq:partfunc}
\end{align}
Such pairwise graphical models provide succinct description for complicated joint distributions. However, the key challenge in utilizing them (e.g. for inference) arises in estimating the partition function $Z(\btheta)$. In this work, our interest is in computing logarithm of $Z(\btheta)$, precisely 
\begin{align}
    \Phi(\btheta) & = \log Z(\btheta)~=~\log\Bigg[\sum_{\bx \in \Xcal^N} \exp\Big(\sum_{e \in E} \theta_e \phi_e(x_e)\Big)\Bigg]. \label{eq:logpart} 
\end{align}

\noindent Computing $Z(\btheta)$ is known to be computationally hard in general, i.e. $\#$P-complete due to relation to counting discrete objects such as independent sets cf. \cite{valiant1979complexity, jerrum1989approximating}. 
Due to reductions from discrete optimization problems to log-partition function computation, approximating 
$\Phi(\btheta)$, even up to a multiplicative error, can be NP-hard cf. \cite{weitz2006counting, wainwright2008graphical,dembo2013factor}. Therefore, the goal is to develop polynomial time (in $N$) approximation method for $\Phi(\btheta)$ with provable guarantees. Specifically, let $\alg$ denote
such an approximation method that takes problem description $\Pcal = (G, \Xcal, (\phi_e)_{e \in E}, \btheta)$ as 
input and produces estimate $\hPhi^{\alg}(\btheta)$ for $\Phi(\btheta)$. Then,
we define approximation ratio associated with $\alg$, $\aprx(G, \alg) \geq 1$ as 
\begin{align}\label{eq:approx.ratio}
    \aprx(G, \alg) & = \sup_{\Pcal} \max\Big(\frac{\Phi(\btheta)}{\hPhi^{\alg}(\btheta)},  \frac{\hPhi^{\alg}(\btheta)}{\Phi(\btheta)} \Big).
\end{align}

\medskip
\noindent {\bf Prior Work.} There is a long literature on computationally efficient approximation methods for the log-partition function with significant progress in the past two decades. We recall some relevant prior works here. 

\noindent A collection of methods, classified as variational approximations, utilize the (Gibbs) variational characterization of the log-partition function when distribution \eqref{eq:pairwise} is viewed as a member of
an exponential family, cf. \cite{georgii2011gibbs, wainwright2008graphical}. Specifically, $\Phi(\btheta)$
can be viewed as a solution of a high-dimensional constrained maximization problem. By solving the problem with additional constraints, one obtains a valid lower bound such as that given by Mean-Field methods. By utilizing
the convexity of $\Phi(\cdot)$ and restricting it to tree-structured sub-graphs of $G$, one obtains a valid upper bound such as that given by the tree-reweighted (TRW) method. By relaxing the constraints and adapting the objective to allow for pairwise pseudo-marginals, one obtains heuristics such as Belief Propagation (BP)
via Bethe approximation \cite{yedidia2001generalized, yedidia2003understanding}. While BP does not 
provide provable upper or lower bounds in general, for graphs with large-girth such as sparse 
random graphs and distributions with spatial decay of correlation, it provides an excellent
approximation cf. \cite{mezard2009information}. The spatial decay of correlation 
property has been further exploited to obtain a deterministic Fully Polynomial Time Approximation
Schemes (FPTAS) for various counting problems, i.e. computing partition functions cf. 
\cite{weitz2006counting, gamarnik2012correlation, bayati2007simple, gamarnik2009sequential}. 
The approximation error of belief propagation has been
studied through connection to loop calculus as well cf. \cite{chertkov2006loop, chandrasekaran2011counting}. 

\noindent In another line of works, graph partitioning based methods have been proposed to provide
Polynomial Time Approximation Schemes (PTAS) for classes of graphs that satisfy certain
graph partitioning properties which includes minor-excluded graphs \cite{jung2006local} or graphs
with polynomial growth \cite{jung2009local}. 

\noindent In summary, despite the progress, the approximation ratio $\aprx(G, \alg)$ for any of the 
known variational approximation methods $\alg$ remains undetermined. 

\medskip
\noindent {\bf Summary of Contributions.} As the main contribution, for a simple variant of tree-reweighted (TRW) method, 
denoted as $\trwp$, we quantify $\aprx(G, \trwp)$ for any $G$. $\trwp$ is described in Section \ref{sec:main}
and produces an estimate of $\Phi(\cdot)$ in polynomial time. Specifically, we establish
\begin{theorem}\label{theorem_main}
For any graph $G$, the approximation ratio of $\trwp$ is such that 
\begin{align}\label{eq:main}
\aprx(G, \trwp) & \leq 1/\sqrt{\kappa(G)} \quad \text{where} \quad    \kappa(G) = \min_{S \subset V} \frac{|S|-1}{|E(S)|},
\end{align}
with $E(S) = E \cap (S \times S)$ for any $S \subset V$. 
\end{theorem}
\noindent The term $\kappa(G)$ captures the proximity of $G$ with respect to the tree structure across 
all of its induced sub-graphs: for $S \subset V$, the induced subgraph $(S, E(S))$ would
have at most $|S|-1$ edges if it were cycle free, but it has $|E(S)|$ edges. Therefore,
the ratio $(|S|-1)/|E(S)|$ measures how far it is from a tree. It is equal to 1 for a connected 
tree and $2/|S|$ for the complete graph. The minimum over all possible $S \subset V$ of this
ratio captures how far $G$ is from a tree structure.

\medskip 
\noindent Using this characterization, we provide bounds on $\aprx(G, \trwp)$ in terms of various simpler graph properties in Section \ref{ssec:kappa.eval}.  Specifically, we show that for any graph with maximum average degree $d \geq 1$, $\aprx(G, \trwp) \leq \sqrt{(d+1)/2}$. And for graphs with girth (i.e. length of shortest cycle) $g > 3$, $\aprx(G, \trwp) \leq \sqrt{\frac{1+N^{2/(g-3)}}{2(1-1/g)}}$, which implies $\aprx(G, \trwp) \leq \big(1 + \frac{1}{2\beta}+o_{\beta \rightarrow \infty}(\frac{1}{\beta})\big)$ if $g \geq \beta \log N$. 
This means that for {\em any} $G$ with large ($\gg \log N$) girth, 
$\aprx(G, \trwp) \approx 1$.

\medskip
\noindent In general, we establish that $\kappa(G)$ can be evaluated in polynomial time for any graph $G$ by solving an appropriate linear program on the spanning tree polytope.
This is explained in Section \ref{sec:kappa}.

\medskip
\noindent The tree-reweighted variant $\trwp$ uses a linear solver over the tree polytope of $G$, which can be hard to implement in practice. With an eye towards near linear-time (in $|E|$) computation, a variant that instead of optimizing over the tree polytope simply considers a feasible point that 
corresponds to the uniform distribution over spanning trees of $G$. Using the near-linear time sampling
of spanning tree from \cite{schild2018almost}, we provide a randomized approximation method for $\Phi(\btheta)$. With high probability, its approximation
ratio $\aprx(G)$ is bounded above by $1/\sqrt{\min_{e \in E} r_e}$ where 
$r_e \geq 0$ is the effective resistance of $e \in E$ for the graph $G=(V,E)$ (see \eqref{eq:eff.res} for
precise definition). While in general, this provides a weaker approximation guarantee, for graphs with degree bounded by $d$ it leads to a similar guarantee of
$\aprx(G) \leq \sqrt{(d+1)/2}$. 

\medskip 
\noindent We show that the results based on graph partitioning cf. \cite{jung2006local, jung2009local} can be 
recovered as a natural extension of the variant of TRW introduced in this work by allowing for general graphs with bounded tree-width beyond trees. 

\medskip\noindent 
We take note of the fact that though results discussed in this work are primarily for the variant of TRW described in Section \ref{sec:main}, as an immediate consequence of our results, $\aprx(G, \trw) \leq 1/\kappa(G)$, i.e. it is bounded by the square of that derived in Theorem \ref{theorem_main}. As discussed in Section \ref{sec:conc}, understanding the tightness of this characterization especially for $\trw$
remains an important open direction.

\medskip \noindent {\bf Outline of Paper.} In Section \ref{sec:background}, we provide some preliminaries
and recall the tree-reweighted (TRW) method. In Section \ref{sec:main}, we provide a modification
of TRW and characterize its approximation guarantee. In Section \ref{sec:kappa}, we provide a linear optimization
characterization of the approximation guarantee which leads to the proof of Theorem \ref{theorem_main}. We also discuss
implications of Theorem \ref{theorem_main} for various classes of graphs. In Section \ref{sec:uniform}, 
we present a near linear-time variant based on sampling from the uniform distribution of spanning
trees over $G$. We derive approximation guarantees for the resulting method in terms of the effective resistance
of the graph and derive its implications. In Section \ref{sec:extension}, we discuss connection with graph
partitioning methods by extending the modified TRW of Section \ref{sec:main} to allow for bounded tree-width
subgraphs beyond trees. We argue how results of \cite{jung2006local, jung2009local} follow naturally. Section \ref{sec:conc} discusses directions for future work.

\section{Preliminaries and Background}\label{sec:background}

\subsection{Variational Characterization, Mean-Field Approximation and Belief Propagation}

We start by recalling the variational characterization of the log-partition function $\Phi(\cdot)$. Let 
$\Pcal(\Xcal^N)$ denote the space of all probability distributions over $\Xcal^N$. Then, the Gibbs
variational characterization states that 
\begin{align}\label{eq_variational}
    \Phi(\btheta) & = \sup_{q \in \Pcal(\Xcal^N)} \Ebb_{\x \sim q}\left(\sum_e \theta_e \phi_e(\x_e) \right) + H(q), 
\end{align}
where $H(q) = -\Ebb_{\x \sim q}(\log(q(\x)))$ is the entropy of $q$. While computationally \eqref{eq_variational} does not provide tractable solution for evaluating $\Phi(\cdot)$, it provides a
framework to develop approximation methods that we refer to as {\em variational approximations}. 

\noindent As mentioned earlier, the classical mean-field (MF) consists in relaxing $\Pcal(\Xcal^N)$ to the space of
independent distributions over $\Xcal^N$ denoted as $\Ical(\Xcal^N)\subset \Pcal(\Xcal^N)$. By restricting
optimization in \eqref{eq_variational} to $\Ical(\Xcal^N)$, one obtains a lower bound on $\Phi(\btheta)$. 


\noindent It turns out that  \eqref{eq_variational} can be solved efficiently for tree-structured graph. Specifically, if $G$ is a connected tree, i.e. $G$ is connected with no cycle, then any distribution
satisfying \eqref{eq:pairwise} can be re-parametrized as 
\begin{align}\label{eq:tree-param}
    \Pb(\bx; \btheta) & = \prod_{u \in V} \Pb_{X_u}(x_u) \prod_{(u, v) \in E} \frac{\Pb_{X_u, X_v}(x_u, x_v)}{\Pb_{X_u}(x_u) \Pb_{X_v}(x_v)}.
\end{align}
In the expression above, $\Pb_{X_u}(\cdot)$ denotes the marginal distribution of $X_u, u \in V$ and 
$\Pb_{X_u, X_v}(\cdot, \cdot)$ denotes the pairwise marginal distribution of $(X_u, X_v)$ for 
any edge $e = (u, v) \in E$. The Belief Propagation (or sum-product) algorithm can compute
these marginal distributions efficiently for tree graphs using only knowledge of $\btheta$ and 
$\phi_e, e \in E$. It utilizes $O(|\Xcal|^2 N)$ computation time. Therefore, $Z(\btheta)$ and hence $\Phi(\btheta)$ can be computed
for tree graphs using $O(|\Xcal|^2 N)$ computations. 

\noindent Indeed, the re-parametrization of the form \eqref{eq:tree-param} was a basis for the
Belief Propagation (BP) algorithm for generic graphical models and also led to the so called
Bethe Approximation of \eqref{eq_variational}, cf. \cite{yedidia2001generalized}. However, it does not
result in a provably upper or lower bound in general (with few exceptions). 

\noindent To obtain an upper bound on  $\Phi(\cdot)$, its convexity was exploited in 
\cite{wainwright2005new} along with the fact that \eqref{eq_variational} is solvable efficiently for tree-structured graph. This resulted into tree-reweighted (TRW) algorithm which we describe next. \\


\subsection{Tree-Reweighted (TRW): An Upper Bound on $\Phi(\cdot)$}

Recall that a spanning tree $T$ is a subgraph of $G=(V,E)$ that contains all vertices and a subset of the edges so that $T$ does not have a cycle. Let $\Tcal(G)$ be the set of all spanning trees of $G$. 
We shall denote a distribution on $\Tcal(G)$ as $\brho = (\rho^T)_{T \in \Tcal(G)}$ where
$\rho^T \geq 0$ for all $T \in \Tcal(G)$ and $\sum_{T \in \Tcal(G)} \rho^T = 1$. The space
of all distributions on $\Tcal(G)$ is denoted by $\Pcal(\Tcal(G))$. For simplicity, we shall
drop the notation of $G$ at times when it is clear from the context and denote it simply as 
$\Pcal(\Tcal)$. A distribution $\brho \in \Pcal(\Tcal)$ induces for all edge $e\in E$ an edge probability $\rho_e$ that this edge will appear in a tree selected from $\brho$,
\begin{align}\label{eq:rho.e}
    \rho_e & = \Pb_{\T \sim \brho}\big(e \in \T) ~=~ \sum_{T \in \Tcal(G)} \rho^T \one(e \in T). 
\end{align}
Note that in the above, we have abused notation using $T$ as a spanning tree as well as the set of edges constituting it. We shall continue using this notation since the all spanning trees have the same set of vertices and only the edges differ (among subsets of $E$). Also note another convenient abuse of notation: given $\brho$, 
$\rho^T$ denotes probability of $T \in \Tcal(G)$ while $\rho_e$ is the probability of edge $e \in E$ being
present in a random tree. Note that because all spanning trees have $N-1$ edges, edge probabilities satisfy $\sum_{e\in E}\rho_e = N-1$. Given $\brho \in \Pcal(\Tcal(G))$, we now define $\kappa(G,{\brho})$ as 
\begin{align}\label{eq:kappa.rho}
    \kappa(G,\brho) & = \min_{e \in E} \rho_e.
\end{align}

\noindent For any $\btheta \in \Rbbp^{|E|}$, define its support as $s(\btheta) = \{ e \in E: \theta_e \neq 0\}$. 
Given a spanning tree $T \in \Tcal(G)$, let $\theta^T \in \Rbbp^{|E|}$ be such that $s(\theta^T) \subset T$ and let
$\brho \in \Pcal(\Tcal)$ along with $(\theta^T)_{T \in \Tcal}$ be such that $\sum_{T \in \Tcal} \rho^T \theta^T = \theta$.
That is, $\Ebb_{\T \sim \brho}\big[\btheta^{\T}\big] = \btheta$. Therefore, we can write 
\begin{align}\label{eq:trw.1}
    \Phi(\btheta) & = \Phi\big(\Ebb_{\T \sim \brho}\big[\theta^{\T}\big]\big).
\end{align}
It has been established that $\Phi: \Rbbp^{|E|} \to \Rbb$ is a convex function because its Hessian is a covariance matrix and hence is positive semidefinite (see \cite{wainwright2005new}). From Jensen's inequality applied to \eqref{eq:trw.1} it follows that 
\begin{align}\label{eq:trw.3}
    \Phi(\btheta) & \leq \Ebb_{\T \sim \brho}\big[ \Phi(\theta^{\T}) \big] ~=~ \sum_{T \in \Tcal} \rho^T \Phi(\theta^T).
\end{align} 
Since the upper bound \eqref{eq:trw.3} holds for any $\brho \in \Pcal(\Tcal)$ and $(\theta^T)_{T\in \Tcal}$ such that $\sum_{T \in \Tcal}\rho^T\theta^{T} = \btheta$ we can optimize on these two parameters to
obtain the tree-reweighted upper bound
\begin{align}\label{eq:trw.4}
    \Phi(\btheta) & \leq \inf_{\sum_{T \in \Tcal}\rho^T\theta^{T} = \btheta}  \left(\sum_{T \in \Tcal} \rho^T \Phi(\theta^T)  \right)~\equiv~U^\trw(\btheta).
\end{align} 
As established in \cite{wainwright2005new}, this seemingly complicated bound, $U^\trw(\btheta)$,
can be computed via an iterative {\em tree-reweighted message-passing} algorithm through the dual of the above 
optimization problem. While this is a valid upper bound, how tight the upper bound is for a given graphical 
model is not quantified in the literature. And this is precisely the primary contribution of this work.

\section{Algorithm and Approximation Guarantee}\label{sec:main}

\noindent{\bf Modified Tree-Reweighted: $\trwp$.} We describe a simple variant of $\trw$ that enables us to bound the approximation
ratio of the estimation of $\Phi$ using properties of $G$. We start with some useful notations. 
Given $\btheta = (\theta_e)_{e \in E} \in \Rbbp^{|E|}$, $\brho \in \Pcal(\Tcal(G))$ 
and a spanning tree $T \in \Tcal(G)$ of graph $G$, define ``projection'' operations
\begin{align}\label{eq:proj.op}
    \Pi^T : \Rbbp^{|E|}\to \Rbbp^{|E|} \quad & \text{where}\quad \Pi^T(\btheta) = \big(\one(e \in T) \theta_e\big)_{e \in E} \nonumber \\ 
    \Pi^T_{\brho} : \Rbbp^{|E|}\to \Rbbp^{|E|}  \quad &\text{where} \quad \Pi^T_{\brho}(\btheta) = 
    \big(\one(e \in T) \theta_e/\rho_e\big)_{e \in E}.
\end{align}
\noindent With these notations, for a given $\brho \in \Pcal(\Tcal(G))$ define
\begin{align}
    L_{\brho}(\btheta) & = \Ebb_{\T \sim \brho}(\Phi(\Pi^{\T}(\btheta))) ~=~ \sum_{T \in \Tcal(G)} \rho^T \Phi(\Pi^T(\btheta)),  \label{eq:est.lb.0} \\
    U_{\brho}(\btheta) & = \Ebb_{\T \sim \brho}(\Phi(\Pi^{\T}_{\brho}(\btheta)))
    ~=~ \sum_{T \in \Tcal(G)} \rho^T \Phi(\Pi^T_{\brho}(\btheta)). \label{eq:est.ub.0}
\end{align}
For a given $\brho \in \Pcal(\Tcal(G))$, one obtains an estimate of $\Phi(\btheta)$ as follows: 
\begin{align}\label{eq:trwp.0}
    \hPhi_{\brho}(\btheta) & = \sqrt{L_{\brho}(\btheta) U_{\brho}(\btheta)}.
\end{align}
$\trwp$ outputs $\hPhi_{\brhos(G)}(\btheta)$ where $\brhos(G)\in\Pcal(\Tcal(G))$ is defined as:
\begin{align}\label{eq:rho.star}
    \brhos(G) & \in ~\argmax_{\brho \in \Pcal(\Tcal(G))} \kappa(G,\brho) \quad\text{with}\quad \kappa(G,\brho) = \min_{e \in E}\rho_e.
\end{align}
\noindent {\bf Interpretation of $\brho^*(G)$.} The distribution $\brho^*(G)\in \Pcal(\Tcal(G))$ can be viewed as a ``balanced cover" of $G$ by a convex combination of its spanning trees (see Appendix \ref{appendix:balanced_covering} for more intuition). The problem of computing $\rho^*(G)$ in polynomial time and characterizing $\kappa(G) = \kappa(G,\brho^*(G))$ is addressed in Section \ref{sec:kappa}. The lemma below quantifies the approximation ratio for $\trwp$. Its proof is in
Appendix \ref{sec:proof_lemma_approx}.
\begin{lemma}\label{lemma_approx} 
Given $\btheta \in \Rbbp^{|E|}$, $\trwp$ produces $\hPhi_{\brhos}(\btheta)$ with 
$\brhos = \brhos(G)$ as defined in \eqref{eq:rho.star}. Then, 
\begin{align}
    \aprx(G, \trwp) & \leq \frac{1}{\sqrt{\kappa(G)}} \quad \text{where} \quad \kappa(G) = \max_{\brho \in \Pcal(\Tcal(G))} \big(\min_{e \in E}\rho_e\big). 
\end{align}
\end{lemma}

\section{$\kappa(G)$: Efficient computation, characterization}\label{sec:kappa}

Lemma \ref{lemma_approx} establishes the approximation guarantee
for $\trwp$ as claimed in Theorem \ref{theorem_main} with caveat
that it is in terms of $\kappa(G) = \max_{\brho \in \Pcal(\Tcal(G))} \big(\min_{e \in E}\rho_e\big)$ while Theorem \ref{theorem_main} relates it to a structural property of the graph defined in 
\eqref{eq:main}. In this section, we shall establish 
this identity which will allow us to bound 
$\kappa(G)$ for certain classes of graphs and obtain meaningful intuitions. In the 
process, we will argue that $\brho^*(G)$ can be computed in polynomial time for
any graph $G$.


\subsection{Computing $\brhos(G)$ and $\kappa(G)$ efficiently}\label{ssec:kappa.compute}

\noindent{\em Spanning Tree Polytope.} We will use a notion of spanning tree polytope for a given graph $G$. Recall
that $\Tcal(G)$ is the set of all spanning trees of $G$. For any tree $T \in \Tcal(G)$, we shall utilize
the notation of $\chi^T = [\chi^T_e] \in \{0,1\}^{E}$ to represent the characteristic vector of the tree $T$
defined such that
\begin{align}
    \forall e\in E: \chi^T_e & = {\mathbf 1}(e \in T).
\end{align}
Given this notation, we define the polytope of spanning trees of $G$, denoted $\Ptree(G)$, as the convex hull of their characteristic vectors. That is,
\begin{align}
    \Ptree(G) & = \big\{ \bv \in [0,1]^{E}: \bv = \sum_{T \in \Tcal(G)} \rho^T \chi^T, ~\sum_{T \in \Tcal(G)} \rho^T = 1, ~\rho_T \geq 0, ~\forall T \in \Tcal(G)  \big\}. 
\end{align}
The weights $(\rho^T)_{T \in \Tcal(G)}$ can be viewed as probability distribution on $\Tcal(G)$, i.e. an element of
$\Pcal(\Tcal(G))$. Therefore $\bv = \sum_{T \in \Tcal(G)} \rho^T \chi^T$ corresponds to a vector representing the probabilities that edges in $E$ will be present in in $\T \sim \brho = (\rho^T)$, i.e. $\bv = \Ebb_{\T \sim \brho}[\one(e \in \T)]$. That is, $\bv = (\rho_e)_{e\in E}$ as defined in \eqref{eq:rho.e}.
Therefore, we shall abuse notation and write
\begin{align}
    \Ptree(G) & = \big\{ (\rho_e)_{e\in E} \mid (\rho^T)_{T\in\Tcal(G)} \in \Pcal(\Tcal(G))\big\}.
\end{align}
\cite{edmonds1971matroids} gave the following characterization of the spanning tree polytope:
\begin{equation}
    \Ptree(G) = \left\{ (v_e)_{e \in E}\in \Rbb_+^E \smiddle| \begin{split}
            \forall S \subset E : v(E(S)) &\leq |S|-1\\
            v(E) &= |V|-1
        \end{split}\right\},
\end{equation}
where $v(E(S)) = \sum_{e\in E(S)}v_e$.\\

\noindent{\em Efficient Separation Oracle.} A polytope $\sf{P} \subset \Rbb^n$, defined through a set of linear constraints, 
is said to have a separation oracle if there exists a polynomial time algorithm in $n$ which for any 
$x\in \Rbb^n$ can determine whether $x \in \sf{P}$ or not; and output a violated constraint if $x \notin \sf{P}$. 
Edmond's characterization of the spanning tree polytope, though it has an exponential number of constraints, 
admits an efficient separation oracle. Such an efficient separation oracle is defined explicitly via a min-cut reduction, see 
\cite[Chapter 4.1]{lau2011iterative}.\\

\noindent{\em Complexity of Linear Programming.} Consider a linear program where the goal is to find the minimum of a 
linear objective function over a polytope defined by finitely many linear constraints. Such a linear program can be
solved in polynomial time (in size of problem description) via the Ellipsoid method if the polytope admits 
an efficient separation oracle, see \cite[Theorem 8.5]{bertsimas1997introduction} for example. Given that
the spanning tree polytope has an efficient separation oracle, optimizing a linear objective over it can be solved efficiently.
Of course, due to the structure of the trees, a greedy algorithm like that of Kruskal's may be a lot more direct
for solving such a linear program. Having said that, the benefit of efficient separation oracle becomes apparent 
as soon as we consider additional linear constraints beyond those described in $\Ptree(G)$. Indeed, such approaches
have found utility in solving other problems, like solving bounded-degree maximum-spanning-tree 
relaxations like in \cite{goemans2006minimum}.  \\

\noindent{\em Augmented Spanning Tree Polytope.} We consider a reformulation of the max-min problem in \eqref{eq:rho.star}. To that end
consider the following augmented spanning tree polytope:
\begin{equation}
    \Ptree_{\min}(G) = \left\{ (z,(v_e)_{e \in E})\in \Rbb\times\Rbb_+^{|E|} \smiddle| \begin{split}
            \forall e\in E: z&\leq v_e\\
            \forall S \subset E : v(E(S)) &\leq |S|-1\\
            v(E) &= |V|-1
        \end{split}\right\}.
\end{equation}
With this notation, we can re-write $\kappa(G)$ as per \eqref{eq:rho.star} as 
\begin{align}
    \kappa(G) & = \max_{(v_e)_{e\in E} \in \Ptree} \{\min_{e\in E}{v_e}\} = \max_{(z,(v_e)_{e\in E}) \in \Ptree_{\min}} z.\label{eq_linear_formulation}
\end{align}
Next, we argue that $\Ptree_{\min}$ admits an efficient separation oracle as follows. 
The separation oracle for $\Ptree_{\min}$ takes $(z,(v_e)_{e\in E})$ as input. 
It first checks that all $|E|$ constraints of the form $z\leq v_e$ are satisfied. 
If one is not satisfied, then the oracle outputs this constraint. If all constraints are satisfied, 
the algorithm runs the separation oracle of $\Ptree$ on $(v_e)_{e\in E}$ and reproduces its output. Since $|E| \leq N^2$ and $\Ptree$ has an efficient separation oracle, this leads to polynomial time
separation oracle for $\Ptree_{\min}$. \\

\noindent{\em Efficient computation of $\brhos(G)$ and $\kappa(G)$.}  
From the linear program formulation \eqref{eq_linear_formulation} and from the efficient 
separation oracle as defined above, we can compute $\kappa(G)$ in polynomial 
time using the Ellipsoid algorithm. Note that this does not directly provides $\brhos(G) \in \Pcal(\Tcal(G))$
since the representation in $\Ptree$ corresponds to the edge probabilities $(\rhos(G)_e)_{e \in E}$. 
However, $(\rhos(G)_e)_{e \in E}$ is a convex combination of extreme points of $\Ptree$, which
correspond to the spanning trees of $G$. Since $\Ptree$ has efficient separation oracle, 
we can recover a decomposition of $(\rhos(G)_e)_{e \in E}$ in terms of convex combination of characteristic vectors weighted by $(\rhos(G)^T)_{T \in \Tcal(G)}$ and such that at most $|E|$ of these weights are strictly positive, see details in \cite[Theorem 3.9]{grotschel1981ellipsoid}. 

\subsection{Characterizing $\kappa(G)$}\label{ssec:kappa.char}

We wish to establish that
\begin{align}
    \kappa(G) & = \max_{(v_e)_{e\in E} \in \Ptree}\{\min_{e\in E}{v_e}\} = \min_{S \subset V}\frac{|S|-1}{|E(S)|}.
\end{align}

\medskip
\noindent{\em Upper bound: $\kappa(G) \leq \frac{|S|-1}{|E(S)|}$.}
The upper bound is immediately given by Edmond's characterisation of the spanning tree polytope. 
For any $(\rho_e)_{e \in E} \in \Ptree$ and any $S \subset V$:
\begin{align}
    |E(S)|\big(\min_{e\in E} \rho_e\big) &\leq \sum_{e\in E(S)} \rho_e~=~\rho(E(S))~\leq~ |S|-1.
\end{align} 
That is, for any $\brho \in \Pcal(\Tcal(G))$
\begin{align}
    \kappa(G,\brho) & \leq \min_{S \subset V} \frac{|S|-1}{|E(S)|}.
\end{align}
And hence it holds for $\brhos(G)$ as well. 

\medskip
\noindent {\em Lower bound: $\kappa(G) \geq \frac{|S|-1}{|E(S)|}$.}
To establish the lower bound, we need a few additional results. To start with, we define
a dual of the optimization problem \eqref{eq_linear_formulation} to characterize $\kappa(G)$. By strong duality it follows that
\begin{align}\label{eq:primal.dual}
    \kappa(G)
    = \max_{\brho \in \Pcal(\Tcal)}\min_{e \in E} \sum_{T\in \Tcal}\one(e\in T)\rho^T 
    = \min_{\bw \in \Pcal(E)} \max_{T \in \Tcal}\sum_{e \in E}\one(e\in T) w_e,
\end{align}
where $\Pcal(E) = \{ \bw = (w_e)_{e \in E} : \sum_{e \in E} w_e = 1, w_e \geq 0 ~\forall~ e\in E\}$. Table \ref{tab:my_label} provides the precise primal and dual formulation associated
with $\kappa(G)$ justifying \eqref{eq:primal.dual}. More intuition and examples are given in Appendix \ref{appendix:balanced_covering}.
\begin{table}[]
    \centering
\begin{tabular}{ccc}
     & \primal & \dual  \\
     Objective & max $z$ & min $y$
     \\
     \\
     Variables / Constraints & \begin{tabular}{rl}
          $z$& $\in \Rbb$  \\
          $\forall T\in \Tcal: \rho_T$&  $\in \Rbb_+$
     \end{tabular}  & 
     \begin{tabular}{rl}
          $\sum_{e \in E}w_e$&$=1$  \\
          $\forall T \in \Tcal: y - \sum_{e\in T}w_e$& $\geq 0$
     \end{tabular}
     \\
     \\
     Constraints / Variables &
     \begin{tabular}{rl}
          $\sum_{T\in \Tcal} \rho_T$& $=1$  \\
          $\forall e \in E: \sum_{T \ni e}\rho_T -z$&  $\geq 0$
     \end{tabular}  & 
     \begin{tabular}{rl}
          $y$&$\in \Rbb$  \\
          $\forall e\in E : w_e$& $\in \Rbb_+$
     \end{tabular}
\end{tabular}
    \caption{The primal (cf. \eqref{eq_linear_formulation}) and dual formulation of $\kappa(G)$.}
    \label{tab:my_label}
\end{table}
We state the following Lemma characterizing an optimal solution of \dual, whose proof is in 
Appendix \ref{proof_appendix_constant_support}. 
\begin{lemma}\label{lemma_constant_support}
There exists an optimal solution of \dual, 
$$\bws \in \argmin_{\bw\in \Pcal(E)} \max_{T \in \Tcal}\sum_{e \in E}\one(e\in T)w_e,$$  
such that all non-zero components of $\bws$ take the identical values.
\end{lemma}
%
%
%

\noindent As per Lemma \ref{lemma_constant_support}, consider an optimal solution $\bws$ of \dual, that assigns 
constant value to a subset $F \subset E$ edges and $0$ to edges $E \backslash F$. It follows that $\ws_e = \frac{1}{|F|}$ for $e \in F$ and $\ws_e = 0$ for $e \in E \backslash F$. Let $V(F) \subset V$ be set of
all vertices corresponding to the end points of edges in $F$ making a subgraph $(V(F), F)$ of $G$. 
Let $c(F) \geq 1$ denote the number of connected components of $(V(F), F)$. Per \dual, given $\bws$, 
$\kappa(G)$ equals the weight of the maximum weight spanning tree in $G$ with edges assigned weights as per $\bws$.
Such a maximum weight spanning tree must select as many edges as possible from $F$: it can select at most $|V(F)|-c(F)$ such edges and any each such edge has weight $1/|F|$ whereas the other selected edges have weight $0$. Thus, the total weight of such a
maximum weight spanning tree is $(|V(F)|-c(F))/|F|$. This gives us an equivalent characterization
for $\kappa(G)$ as 
\begin{align}\label{eq:kappa.1}
\kappa(G) & = \min_{F \subset E} \frac{|V(F)|-c(F)}{|F|}.
\end{align}
Now we state a Lemma, whose proof is in Appendix \ref{proof_lemma_subgraph_to_subset}, that concludes on Theorem \ref{theorem_main}.
\begin{lemma}\label{lemma_subgraph_to_subset} 
For any graph $G$, 
\begin{equation}
    \min_{S \subset V} \frac{|S|-1}{|E(S)|} = \min_{F\subset E} \frac{|V(F)|-c(F)}{|F|}.
\end{equation}
\end{lemma}


\subsection{Evaluating $\kappa(G)$ For Certain Graphs}\label{ssec:kappa.eval}

As established in Section \ref{ssec:kappa.compute}, $\kappa(G)$ can be computed
in polynomial time for any $G$. Here, we attempt to obtain a (lower) bound on $\kappa(G)$ in terms of simple
graph properties. To that end, we obtain the following for graphs with bounded maximum average degree and for graphs with bounded girth. 
\begin{lemma}\label{lem:special.1} 
Consider a graph $G = (V, E)$. If $G$ has maximum average degree $\bar{d} = \max_{S \subset V} \frac{2|E(S)|}{|S|}$, 
\begin{align}
     \frac{2}{\bar{d}+1}\leq \kappa(G).
\end{align}
If $G$ has girth (length of its shortest cycle) $g > 3$,
\begin{align}
    \frac{2}{1+N^{\frac{2}{g-3}}}(1-1/g) \leq \kappa(G).
\end{align}
\end{lemma}
The proof of Lemma \ref{lem:special.1} is presented in Appendix \ref{proof_lemma_bound_kappa}. As an immediate consequence, for $g = \beta \log N$,
\begin{align}
    \kappa(G) = 1-\frac{1}{\beta}+o_{\beta\to\infty}(\frac{1}{\beta}) \quad \text{therefore}\quad  \aprx(G, \trwp) & \leq \frac{1}{\sqrt{\kappa(G)}} =1+\frac{1}{2\beta}+o_{\beta\to\infty}(\frac{1}{\beta}).
\end{align}


\section{A Near Linear-Time Variant of $\trwp$}\label{sec:uniform}

\subsection{Algorithm}

\noindent $\trwp$ requires finding $\brhos(G)$. As discussed in Section \ref{sec:kappa}, it can be computed efficiently. 
However it can be cumbersome and having near-linear (in $|E|$) time variant can be more attractive in practice. With this
as a motivation, we propose utilizing the uniform distribution on $\Tcal(G)$, denoted as $\Ucal(G) \equiv \mathcal{U}(\Tcal(G))$,  in place of $\brhos(G)$ in $\trwp$ . The challenge is it has very large support and hence it is
difficult to compute $L_{\Ucal}(\btheta), U_{\Ucal}(\btheta)$. But, both of these quantities are averages, with respect to
$\Ucal$, of a certain functional. And it is feasible to sample spanning trees uniformly at random for any $G$ in near-linear 
time. Therefore, we can draw $n$ samples from the distribution $\bu$ and consider the empirical distribution $\hbu^n$ to compute estimates $L_{\hbu^n}(\btheta), U_{\hbu^n}(\btheta)$ 
with few samples. This is precisely the algorithm.

\medskip
\noindent To that end,
consider $n$ trees $\T_1,\dots, \T_{n}$ sampled uniformly at random from $\Tcal(G)$. Compute 
\begin{align}
    \hu^n_e & = \frac{1}{n}\sum_{i=1}^{n} \one(e \in \T_i), \quad L_{\hbu^n}(\btheta) ~=~ \frac1n \sum_{i=1}^n \Phi(\Pi^{\T_i}(\btheta)),
    \quad U_{\hbu^n}(\btheta)  = \frac1n \sum_{i=1}^n \Phi(\Pi^{\T_i}_{\hbu^n}(\btheta)), \label{eq.approx.unif}
\end{align}
where $\hbu^n = (\hu^n_e)_{e \in E}$.
Given this, produce the estimate, 
\begin{align}
    \hPhi_{\hbu^n}(\btheta) & = \sqrt{L_{\hbu^n}(\btheta) U_{\hbu^n}(\btheta)}. 
\end{align}

\subsection{Guarantees}

Given a graph $G$, remember that $\kappa(G,\bu) = \min_{e \in E} u_e$ with
$u_e = \mathbb{P}_{\T \sim \bu}(e\in \T)$. We obtain the following guarantee (proof can be found in Appendix 
\ref{proof_finite_sample}). 

\begin{lemma}\label{lemma_finite_sample}
Given $\epsilon > 0$ and $\delta >0$, for $n\geq O\left(\log(\frac{N}{\delta})\kappa(G,\bu)^{-2}\epsilon^{-2}\right)$ and $\epsilon$ sufficiently small, with probability at least $1-\delta$, 
\begin{align}\label{eq:unif.approx}
\max\Big(\frac{\Phi(\btheta)}{\hPhi_{\hbu^n}(\btheta)}, \frac{\hPhi_{\hbu^n}(\btheta)}{\Phi(\btheta)}\Big)
& \leq \frac{1+\epsilon}{\sqrt{\kappa(G,\bu)}}.
\end{align} 
\end{lemma}

\subsection{Computation Cost}

\noindent To sample trees uniformly at random from $\Tcal(G)$, \cite{schild2018almost} recently proposed a "short-cutting" method that has 
$O(|E|^{1+o(1)})$ runtime. The earliest polynomial
time algorithm has been known since \cite{guenoche1983random}. While we do not recall either of these here, we briefly 
recall algorithm from \cite{broder1989generating} due to its elegance even though it is not the optimal (it has
$O(N|E|)$ run time): (1) starting with any $u \in V$ run a random walk on $G$ until it covers all vertices, 
(2) for every vertex $v \neq u$, select the edge through which $v$ was reached for the first time during the walk, 
and (3) output the $N-1$ edges (which form tree) thus selected. 

\medskip
\noindent Given $n$ such samples, in order to compute $\hPhi_{\hbu^n}(\btheta)$, we have to compute 2$n$ log-partition functions for 
tree structured graph. As noted in Section \ref{sec:background}, each such computation requires $O(N|\Xcal|^2)$ operations. By Lemma \ref{lemma_finite_sample}, we therefore need total of 
$O(|E|^{1+o(1)} + N|\Xcal|^2)\times O(\kappa(G,\bu)^{-2} \epsilon^{-2} \log 1/\epsilon)$ runtime for  
$(1+\epsilon)/\sqrt{\kappa(G,\bu)}$ approximation with probability $1-\epsilon$.

\subsection{$\kappa(G,\bu)$ and Effective Resistance}
Recall that $\kappa(G,\bu) = \min_{e \in E} u_e$ where $u_e = \Ebb_{\T \sim \bu}[\one(e \in \T)]$ is equal to
the so called ``effective resistance'' associated with edge $e \in E$ for the graph $G$. The notion was 
introduced by \cite{klein1993resistance} and has multiple interpretations. We present one here. For 
$e =(s,t) \in E$,  the effective resistance $u_e$ is equal to the amount of electric energy dissipated 
by the network when all edges are seen as electric wire of resistance $R_e = 1$ and a generator guarantees 
a total current flow ($\iota_\text{gen}=1$) from $s$ to $t$. The distribution of the current $\iota$ across the network 
must minimize the dissipated energy while respecting the constraints imposed by Kirchoff's laws 
(also see \cite[Chapter 2]{lyons2017probability}). Below we provide variational characterization of it.
\begin{equation}\label{eq:eff.res}
\forall e = (s,t) \in E: \quad u_e = \min \left\{\sum_{\{u,v\}\in E} \iota(u,v)^2 \smiddle| 
    \begin{aligned}
        \forall \{u,v\} \in E: \iota(u,v) + \iota(v,u) &= 0\\
    \forall u \in V\setminus \{s,t\}: \sum_{v\mid(u,v) \in E} \iota(u,v) &= 0 \\
    \sum_{v\mid(s,v) \in E} \iota(s,v) = \sum_{u\mid(u,t) \in E} \iota(u,t) &= 1
\end{aligned}
    \right\}.
\end{equation}

\noindent
We observe that the minimum effective resistance of the graph, $\kappa(G, \bu) = \min_{e\in E} u_e$ can also be connected to structural properties of the graph (proofs are in \ref{proof_lemma_eff_res}).
\begin{lemma}\label{lemma_eff_res} 
Given $G = (V, E)$, if $G$ has maximum degree $d$,
\begin{equation}
    \frac{2}{d+1}\leq \kappa(G,\bu) 
\end{equation},
and if $G$ has girth at least $g > 3$,
\begin{equation}
    \frac{1}{1+ |E|/(g-1)^2} \leq \kappa(G,\bu) 
\end{equation}
\end{lemma}

\section{Beyond Trees}\label{sec:extension}

This far, we have restricted to approximating $\Phi(\btheta)$ by decomposing 
$\btheta = \Ebb_{\T \sim \brho}[\Pi_{\brho}^T(\btheta)]$ and then using properties of $\Phi$ to produce an approximation 
guarantee. Such arguments would hold even if we decompose $\btheta$ using
subgraphs of $G$ beyond trees. The choice of trees was particularly useful since
they allow for an efficient computation of $\Phi$. In general, graphs with bounded
tree-width lend themselves to efficient computation of $\Phi$, cf. \cite{chandrasekaran2011counting}.

\medskip
\noindent To that end, let $\Tcal_k(G)$ denote the set of all subgraphs of $G$ that have treewidth bounded
by $k \geq 1$. Let $\Pcal(\Tcal_k(G))$ denote the distribution over all such subgraphs. For any 
$H \in \Tcal_k(G)$ and $\brho \in \Pcal(\Tcal_k(G))$, define $\Pi^H(\cdot)$ and $\Pi^H_{\brho}(\cdot)$
similar to that in \eqref{eq:proj.op} in Section \ref{sec:main} and define,
\begin{align}
    L_{\brho}(\btheta) & = \Ebb_{\Hs \sim {\brho}}(\Phi (\Pi^{\Hs} (\btheta))), \quad 
    U_{\brho}(\btheta)  = \Ebb_{\Hs \sim \brho}(\Phi(\Pi_{\brho}^{\Hs}(\btheta))), \quad 
    \hPhi_{\brho}(\btheta)  = \sqrt{L_{\brho}(\btheta)U_{\brho}(\btheta)}.
\end{align}
 Using identical arguments as in Theorem \ref{theorem_main}, it follows that $\hPhi_{\brho}(\btheta)$
 is $1/\sqrt{\kappa(G,\rho)}$-approximation. By optimizing over the choice $\rho^*_k \in \Pcal(\Tcal_k(G))$ one would theoretically attain  
 \begin{align}\label{eq:kappa.k}
     \kappa_k(G) & = \max_{\brho \in \Pcal(\Tcal_k(G))} \min_{e \in E} \rho_e.
 \end{align}

\noindent{\em $(\epsilon,k)$-partitioning.} While such generality is pleasing, the space of $\Pcal(\Tcal_k(G))$ seems too vast and complex to compute $\kappa_k(G)$ or the associated distribution $\rho_k^*(G) \in \argmax_{\brho \in \Pcal(\Tcal_k(G))} \min_{e \in E} \rho_e$ for $k>1$. In \cite{jung2006local, jung2009local} a seemingly different approach
was proposed using graph partitioning. It resulted in an approximation method for a large family of graphs including
minor-excluded graphs and graphs with polynomial growth. It considers the set of $k$-partitions of $G$, $\Prt_k(G) \subset \Tcal_k(G)$ defined as
\begin{equation}
    \Prt_k(G) = \{H = (V,\bigcup_{i=1}^K E(S_i))  \mid (S_i)_{1\leq i \leq k} \text{ is a partition of } V \text{ and } \forall i: |S_i| \leq k\}.
\end{equation}
A distribution $\brho \in \Pcal(\Prt_k(G)) \subset \Pcal(\Tcal_k(G))$
is called an $(\epsilon,k)$-partitioning of $G$ if $\forall e\in E: 1-\epsilon \leq \Ebb_{\Hs \sim \brho}[\one(e \in \Hs)] \leq 1$. The proof of the following result can be found in Appendix \ref{proof_partitionning}.
\begin{theorem}\label{theorem_partitionning} 
If $G$ is such that there exists an $(\epsilon, k)$-partitioning of $G$, $\brho\in \Prt_k(G)$, then
\begin{align}
    \sqrt{1-\epsilon} & \leq \frac{\Phi(\btheta)}{{\hPhi}_\rho(\btheta)}\leq \frac{1}{\sqrt{1-\epsilon}}.
\end{align}
\end{theorem}
We note that$\frac{1}{\sqrt{1-\epsilon}} = 1 + \frac{1}{2}\epsilon + o(\epsilon)$ and hence it 
improves upon the result given in \cite{jung2006local, jung2009local} which achieves a 
$1+\epsilon$ approximation error. 

\section{Relation to hardness of approximation}\label{sec: hardness}

As mentioned in earlier, the task of computing log-partition function $\Phi(\btheta) = \log(Z(\btheta))$ 
is computationally hard. Further finding constant factor approximation to it is hard in general. We
provide details to this effect. 

\medskip
\noindent To that end, let us consider a graph $G$ with $\Xcal = \{-1,1\}$ and potentials such that 
$\forall e = (u,v) : \phi_e(x_e) = \beta\1(x_u\neq x_v)$ where $\beta \in \Rbb^+$. 
The corresponding partition function $Z(\beta)$ satisfies:
\begin{align}
    Z(\beta) &= \sum_{\bx \in \{-1,1\}^n}\exp\left(\beta \sum_{(u,v)\in E}\1(x_u\neq x_v)\right),\\
    &= \sum_{0\leq k \leq |E|}N_k\exp(\beta k),
\end{align}
where $N_k$ denotes the number of $k$-cuts of $G$. This gives the following bound:
\begin{align}
    \beta\maxcut(G)&\leq \Phi(\beta) \leq N\log(2)+\beta\maxcut(G),
\end{align}
where $\maxcut(G)= \max\{k \mid N_k \neq 0\}$ is the cardinality of the maximum cut of $G$. 
As a consequence, for $\beta$ sufficiently large ($\beta = N\log(2)/\epsilon$ for some small $\epsilon>0$) 
observe that any constant factor guarantee on $\hPhi(\beta)$ will yield a similar constant factor guarantee 
on the size of the maximum cut of $G$. The task of approximating the maximum cut of a graph was shown to be 
NP-hard for any constant factor greater than $16/17\approx 0.941$ by \cite{haastad2001some} for general graphs, 
although this ratio was improved under structural assumptions for $G$ (such as bounded degree) 
by \cite{feige2002improved}. Through the above connection, our result provides $1/\kappa(G)$-approximation 
guarantee for any $G$.


\medskip
\noindent The example above generalizes to binary constraint satisfaction problems (CSP). 
For a graph $G = (V,E)$ and an alphabet $\Xcal$, a binary constraint satisfaction problem $\Ccal$ 
is defined by a collection of constraints on the edges of the graph $(C_e)_{e\in E}$, where each constraint 
corresponds to a subset $C_e \subset \Xcal^2$ of ``acceptable" values for the corresponding pair of variables. 
For a given constraint satisfaction problem $\mathcal{C} = (C_e)_{e\in E}$, a quantity of interest is its 
maximum satifiability $\maxcsp(\Ccal)$, which is the highest fraction of satisfied constraints, over all 
possible assignments for $V$. Notice that the maximum cut problem mentioned above is a particular case. 
By the same derivations, we obtain that for the choice of potential functions $\phi_e(x_e)=\beta\1(x_e\in C_e)$ and 
$\beta$ sufficiently large ($\beta \geq N\log(|\Xcal|)/\epsilon$) the corresponding log-partition function 
satisfies:
\begin{equation}
    \Phi(\beta)/\beta -\epsilon \leq \maxcsp(\Ccal) \leq \Phi(\beta)/\beta.
\end{equation}
Maximum constraint satisfaction problems are known to be hard to approximate. A particular case of constraint satisfaction 
problems (where the constraints are of the form $C_e = \{(z,\pi(z))\mid z\in \Xcal\}$ with $\pi$ a permutation of $\Xcal$) 
are called ``unique games" (see \cite{khot2005unique} for precise definition). The unique games conjecture \citep{khot2002power} 
states that no constant factor approximation guarantee is achievable in polynomial time for maximum constraint satisfaction 
problems on unique games in general. That is, one should not expect constant factor approximation for log partition function
computation, even with non-negative potentials as considered in this work, for general graph. Indeed, this work provides such
non-constant factor approximation, $1/\kappa(G)$, for any graph $G$ for tree-reweighted method. The question remains, whether 
such is the tightest possible for tree-reweighted method in general.

\section{Conclusions}\label{sec:conc}

We presented an approach to quantify the approximation ratio of variational methods for estimating the log-partition
function of discrete pairwise graphical models. As the main contribution, we quantified the approximation 
error as a function of the underlying graph properties. In particular, for a variant of the tree-reweighted algorithm, 
for graphs with bounded degree the approximation ratio is a constant factor (function of degree) and for graphs with large ($\gg$ logarithmic) girth, the approximation ratio is close to $1$. The method naturally extends beyond trees unifying prior works on graph partitioning based approach. 

In this work, we restricted the analysis to non-negative valued potentials and edge parameters. If potentials
are bounded, we can transform the general setting into a setting with
non-negative potentials. However, the approximation ratio with respect to this transformed setting
may not translate to that of the original setting. This may be interesting direction for future works.

\acks{This work is supported in parts by projects from NSF and KACST as well as by a Hewlett Packard graduate fellowship. We would like to thank Moïse Blanchard for interesting discussions on duality and three anonymous reviewers for their useful comments. }
\bibliography{references}
\appendix

\section{Balanced Covering of Graphs}\label{appendix:balanced_covering}
\begin{figure}[htp]
    \centering
    \includegraphics[width=10cm]{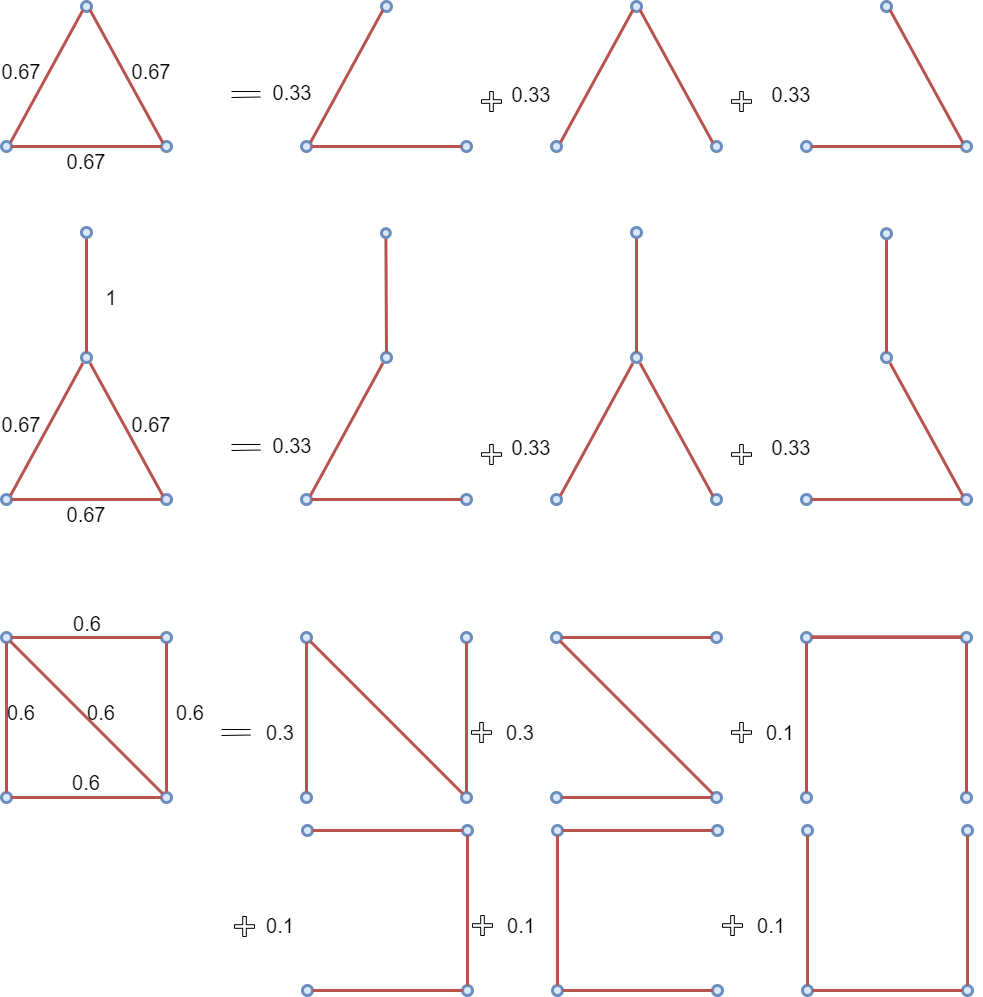}
    \caption{Three examples of balanced coverings of graphs. Notice that for the triangle, the most connected subgraph is the entire graph itself which yields $\kappa(G) = \frac{3-1}{3} = \frac{2}{3}=0.67$ just like for the third example for which $\kappa(G) = \frac{4-1}{5}=\frac{3}{5}=0.6$.}
    \label{fig:balanced_covering_examples}
\end{figure}

\begin{figure}[htp]
    \centering
    \includegraphics[width=15cm, trim={2.5cm 2.5cm 2.5cm 2.5cm},clip]{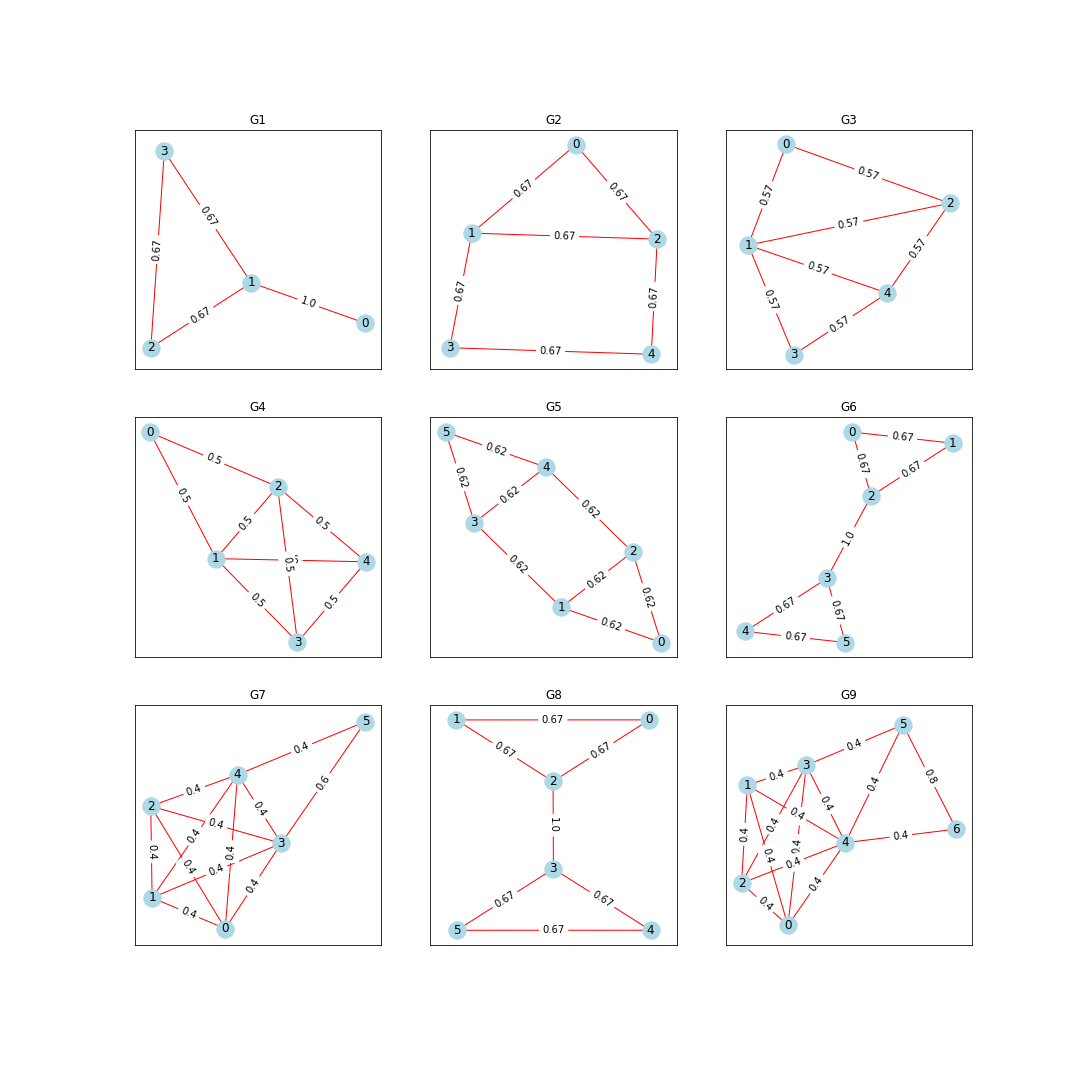}
    \caption{More examples of balanced covering of graphs, obtained with a LP solver. The number assigned to each edge corresponds to $(\rho_e)_{e\in E}$ for a balanced cover $\brho^*(G)$. Note that for $G_7$, some of the edge probability of edge $(3,5)$ could be redistributed to edge $(4,5)$ by symmetry, but this would not improve the optimum. This is also true for $G_9$.}
    \label{fig:balanced_covering_simulations}
\end{figure}

In section \ref{sec:kappa}, we described two linear programming problems $\primal$ and $\dual$ of optimum equal to the main quantity of interest $\kappa(G)$. We recall their formulation given in \eqref{eq:primal.dual} and provide some interpretation.
\begin{align*}
    \primal
    = \max_{\brho \in \Pcal(\Tcal)}\min_{e \in E} \sum_{T\in \Tcal}\one(e\in T)\rho^T 
    = \min_{\bw \in \Pcal(E)} \max_{T \in \Tcal}\sum_{e \in E}\one(e\in T) w_e = \dual.
\end{align*}
The objective of $\primal$ is to obtain a \textit{balanced covering} of G with its spanning trees, that is a distribution $\brho \in \Pcal(\Tcal(G))$ that maximizes the probability of apparition of the least likely edge. We give example of such balanced coverings in Figure \ref{fig:balanced_covering_examples}. Interestingly, $\dual$ also has a nice formulation as a \textit{max-min spanning tree} problem. Given a total budget of $1$, it aims to assign weights to the edges of $G$ such that an adversary removing a maximum spanning tree will receive the smallest reward. It is natural to observe that the densest sub-graphs of $G$ are the key to both these problems. They are key to $\primal$ because they are harder to cover with spanning trees. They are key to $\dual$ because assigning weight to dense regions is a good way to limit the adversary's reward. This is made explicit by the representations in Figure \ref{fig:balanced_covering_simulations} and Figure \ref{fig:optimal_weights}. 

\section{Proof of Lemma \ref{lemma_approx}}\label{sec:proof_lemma_approx}

 We start by observing a few properties of function $\Phi(\cdot)$. 

\noindent{\em Property 1. $\Phi$ is non-decreasing.} For $a, b \in \Rbb^n$ let $a \preceq b$ denote that 
every component of $a$ is less or equal to that of $b$, i.e. $a_i \leq b_i, ~i \in [n]$. 
With this, for $\btheta, \btheta^\prime \in \Rbbp^{|E|}$ such that $\btheta \preceq \btheta^\prime$, it
can be easily verified that
\begin{align}\label{eq:monotonicity}\tag{monotonicity}
    \Phi(\btheta) & \leq \Phi(\btheta^\prime).
\end{align}
Since $\Phi(\bzero) = N \log |\Xcal|$, and $\bzero \preceq \btheta \preceq \btheta^\prime$, we have
\begin{align}
     N\log(|\Xcal|) & \leq \Phi(\btheta) ~\leq~ \Phi(\btheta^\prime).
\end{align}

\noindent{\em Property 2. $\Phi$ is sub-linear.} For $\lambda \geq 1$ and $\btheta \in \Rbbp^{|E|}$, 
\begin{align}\label{submodularity}\tag{sub-linearity}
     \Phi(\lambda \btheta) & \leq \lambda \Phi(\btheta).
\end{align}
The above follows from the fact that for any $\bs = (s_i) \in \Rbbp^n$, 
\begin{align*}
    \big(\sum_{i=1}^n s_i^\lambda\big) & \leq \big(\sum_{i=1}^n s_i\big)^\lambda. 
\end{align*}

\noindent Now consider $\brho \in \Pcal(\Tcal(G))$. For any $T \in \Tcal(G)$ and $\btheta \in \Rbbp^{|E|}$, 
by definition of $\Pi^T$, we have that $\Pi^T(\btheta) \preceq \btheta$. Therefore, using the monotonicity of the log-partition function
it follows that
\begin{align}
    L_{\brho}(\btheta) & = \sum_{T \in \Tcal(G)}\rho^T \Phi(\Pi^T(\btheta)) 
    \leq  \sum_{T \in \Tcal(G)}\rho^T \Phi(\btheta) \leq \Phi(\btheta). \label{eq:trwp.lb.2}
\end{align}
By definition $\btheta = \Ebb_{\T \sim \brho}[\Pi^{\T}_{\brho}(\btheta)]$, and due to convexity of 
$\Phi$, it follows that
\begin{align}\label{eq:trwp.ub.2}
    \Phi(\btheta) & = \Phi\big(\Ebb_{\T \sim \brho}[\Pi^{\T}_{\brho}(\btheta)]\big) ~\leq~\Ebb_{\T \sim \brho}[\Phi(\Pi^{\T}_{\brho}(\btheta))] ~=~U_{\brho}(\btheta).
\end{align}
By definition of $\kappa(G,\brho) = \min_{e \in E} \rho_e$, it follows that 
\begin{align}
    \Pi^T_{\brho}(\btheta) & \leq \frac{1}{\kappa(G,\brho)} \Pi^T(\btheta), ~~\forall ~T\in \Tcal(G).
\end{align}
And, by definition $1/\kappa(G,\brho) \geq 1$. Therefore by \eqref{eq:monotonicity} and \eqref{submodularity}, we have
\begin{align}
    \Phi(\Pi^T_{\brho}(\btheta)) & \leq \Phi\big(\frac{1}{\kappa(G,\brho)} \Pi^T(\btheta)\big) ~\leq~ \frac{1}{\kappa(G,\brho)}\Phi( \Pi^T(\btheta)).
\end{align}
Therefore,
\begin{align}\label{eq:trw.3}
    U_{\brho}(\btheta) & = \sum_{T \in \Tcal(G)} \rho^T \Phi(\Pi^T_{\brho}(\btheta)) ~\leq~\frac{1}{\kappa(G,\brho)} \sum_{T \in \Tcal(G)} \rho^T \Phi(\Pi^T(\btheta))~=~ \frac{1}{\kappa(G,\brho)} L_{\brho}(\btheta). 
\end{align}
As a consequence of \eqref{eq:trwp.lb.2}, \eqref{eq:trwp.ub.2} and \eqref{eq:trw.3} we obtain that
\begin{align}\label{eq:trw.4}
    \Phi(\theta) \leq U_{\brho}(\btheta) & \leq \frac{1}{\kappa(G,\brho)}\Phi(\btheta)  \quad \text{and} \quad \kappa(G,\brho)\Phi(\btheta) \leq L_{\brho}(\btheta) \leq \Phi(\btheta).
\end{align}
From this, it follows that
\begin{align}
    \sqrt{\kappa(G,\brho)}\Phi(\btheta) \leq \sqrt{L_{\brho}(\btheta)U_{\brho}(\btheta)} \leq \frac{1}{\sqrt{\kappa(G,\brho)}}\Phi(\btheta).
\end{align}
Which can be rewritten as
\begin{align}
    \sqrt{\kappa(G,\brho)} \leq \frac{\hPhi_{\brho}(\btheta)}{\Phi(\btheta) } \leq \frac{1}{\sqrt{\kappa(G,\brho)}}.
\end{align}
By optimizing over choice of $\brho = \brhos(G)$, we conclude
that $\aprx(G, \trwp) \leq \frac{1}{\kappa(G)}$.

\section{Proof of Lemma \ref{lemma_constant_support}}\label{proof_appendix_constant_support}

\begin{figure}[htp]
    \centering
    \includegraphics[width=15cm, trim={2.5cm 2.5cm 2.5cm 2.5cm},clip]{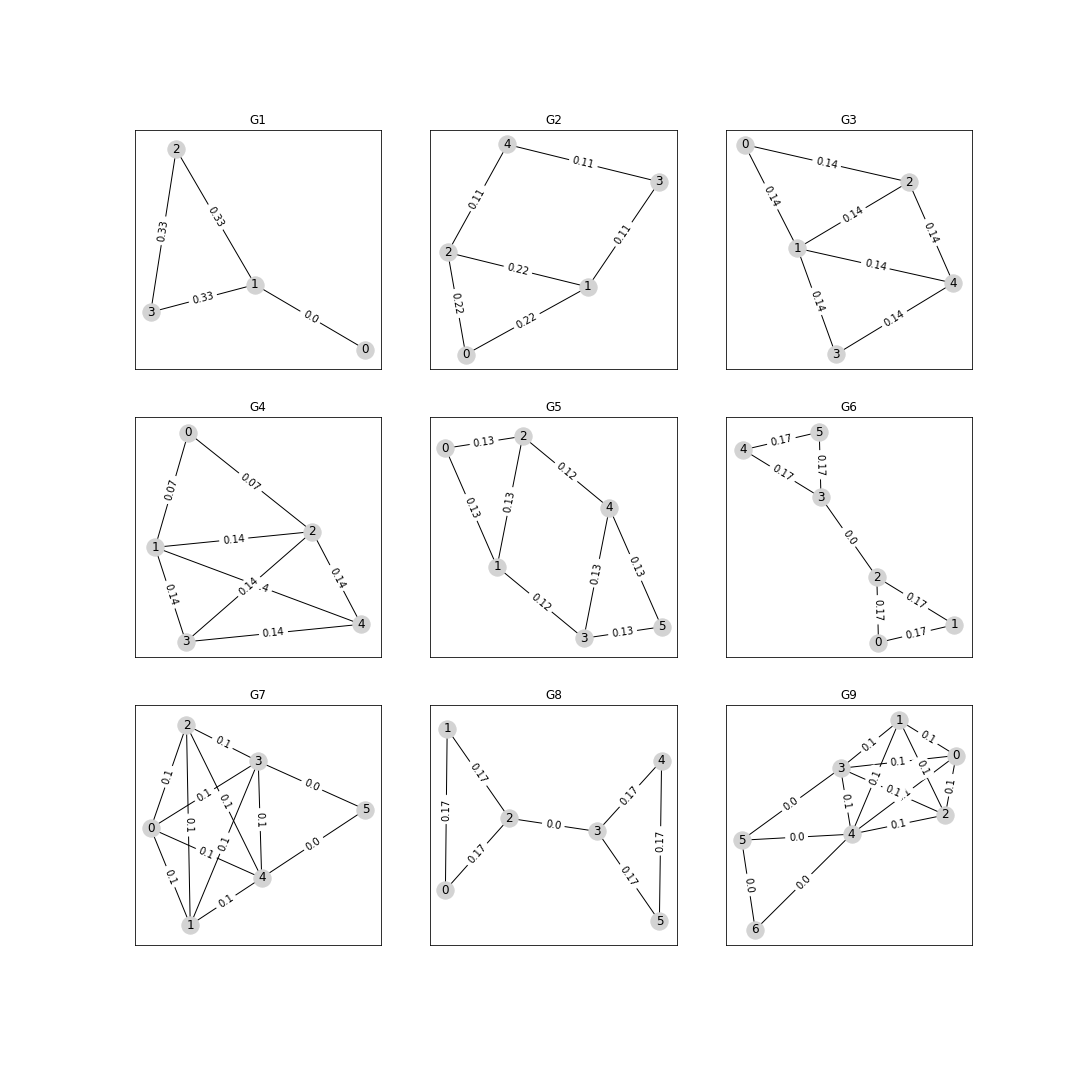}
    \caption{An example of an optimal weight assignment $(w_e)_{e\in E}$ for the problem $\dual$ on nine different graphs. The solution was found by the interior point method using a linear programming solver. Note that for most graphs, the solution reached is already constant on its support. On graphs $G_2$ and $G_4$, note that evening out the weights would not increase the weight of the maximum spanning tree.}
    \label{fig:optimal_weights}
\end{figure}

 (See illustration in Figure \ref{fig:optimal_weights}) For $\bw = (w_e)_{e\in E}$ denote $f(\bw)$ the number of distinct values in its support:
\begin{equation}
f(\bw) = |\{w_e : e \in E, w_e \neq 0\}|.
\end{equation}
To prove the lemma, it suffices to show that there exists an optimal solution of $\dual$ such that $f(\bw) = 1$. We will prove that if $\bw$ is an optimal solution and $f(\bw) > 1 $ then we can build $\bw'$ of similar objective value such that $f(\bw')\leq f(\bw)-1$. By repeating
this till $f(\bw) = 1$ will conclude the proof.\\ 

\noindent Let $\bw$ be an optimal solution with $f(\bw)>1$. 
We consider the edges $e_1, e_2, ..., e_{|E|}$ ordered 
by their weights, i.e. 
\begin{equation}
    w_{e_1} \geq ... \geq w_{e_{|E|}}.
\end{equation}
In what follows, we will make sure that the ordering on the edges never changes, therefore we allow ourselves to write $w_i$ instead of $w_{e_i}$. Now the objective of \dual~ achieved by such an optimal
$\bw$ corresponds to the weight of a maximum weight spanning tree. 
Let us utilize Kruskal's algorithm to find such an maximum weight
spanning tree. Recall that Kruskal's algorithm greedily selects
edges from higher to lower weight as long as they do not create 
a cycle with previously selected edges. We will denote  
$I_T = \{t_1 < ... < t_{N-1}\}$  the indices of the edges
selected by the algorithm to construct tree $T$ and let 
$I_{E\setminus T} = \cup_{k=1}^{N-1} \{s: t_k < s < t_{k+1}\}$
denote the indices of edges not part of $T$ with notation
$t_{N} = |E|+1$. The weight of the maximum spanning tree 
is then $\bw(T) = \sum_{k=1}^{N-1} w_{t_k}$. 
Note that $t_1 = 1$ and $t_2 = 2$ since cycle requires $3$ or more
edges. By definition $w_{j-1} \geq w_j$ for $2\leq j \leq |E|$. 
Now if $w_{j-1} > w_j$ the we claim that $j \in I_T$. This is because for $1\leq k \leq N-1 $ if $(w_{t_k},....,w_{t_{k+1}-1})$ 
are not equal, setting them all to their average 
decreases $w_{t_k}$ strictly while preserving 
$\bw \in \Pcal(E)$ as well as the order on the edges and 
therefore contradicting the optimality of $\bw$ for \dual.
Therefore $\bw$ is piece-wise constant with discontinuities 
only appearing for $j\in I_T$.\\

If $f(\bw)=2$ and all weights are positive, we denote 
$2 \leq k \leq N-1$ such that $w_{t_k-1}>w_{t_{k}}>0$ and we have:
\begin{equation}
    w_{1} = ... = w_{t_{k}-1} > w_{t_{k}} = .... = w_{|E|}.
\end{equation}
In this case, the optimal objective value for \dual~ is equal to $(k-1)w_1 + (N-k)w_{t_k} $. 
To make $\bw$ constant on its support while preserving the order on the weights, there are two possibilities. Either transfer all weight from $(w_{t_{k}}, .... ,w_{|E|})$ to $(w_{1}, ... , w_{t_{k}-1})$ until $(w_{t_{k}}, .... ,w_{|E|})$ reaches zero. The objective  will then
be $w_1 + \frac{|E|-t_k+1}{t_k-1}w_{t_k}$. Or transfer all weight from $(w_{1}, ... , w_{t_{k}-1})$ to $(w_{t_{k}}, .... ,w_{|E|})$ until all weights are equal. The objective will be then $w_{t_k} + \frac{|E|-t_k+1}{t_k-1}(w_1 - w_{t_k})$. Because either $\frac{|E|-t_k+1}{t_k-1}\leq 1$ or $\frac{t_k-1}{|E|-t_k+1}\leq 1 $, one of these transfers does not increase the objective and yields $f(\bw)=1<2$.\\

If $f(\bw)=2$ and some weights are $0$, 
denote $k_0$ the smallest index such that $w_{t_{k_0}} = 0$. The method above still holds when replacing $|E|-t_k+1$ by $t_{k_0}-t_k$.\\

Now suppose $f(\bw) \geq 3$, making sure that the order on the weights is preserved requires extra caution. In addition to $k$ and $k_0$ (if required), we denote $k_1$ the index of the discontinuity that follows $k$. We have: 
\begin{equation}
    ... = w_{t_{k_1}-1} > w_{t_{k_1}} = ... = w_{t_{k}-1} > w_{t_{k}} = ... = w_{t_{k_0}-1} > w_{t_{k_0}} = ...
\end{equation}
In the event when we want to transfer weight from $(w_{t_{k}}, ..., w_{t_{k_0}-1})$ to $(w_{t_{k_1}}, ..., w_{t_{k}-1})$, we must make sure that $(w_{t_{k_1}}, ..., w_{t_{k}-1})$ does not exceed $w_{t_{k_1}-1}$. If $(w_{t_{k_1}}, ..., w_{t_{k}-1})$ attains $w_{t_{k_1}-1}$ the transfer must stop at equality, and one should observe that we have decreased $f(\bw)$ strictly by $1$ because the discontinuity at $w_{t_{k_1}}$ has disappeared and no new discontinuity was created. \\

In summary, we have argued that if $\bw$ is an optimal solution and $f(\bw) > 1 $ then we can build $\bw'$ of same objective value (optimal) and such that $f(\bw')\leq f(\bw)-1$. This completes the proof of Lemma.

\section{Proof of Lemma \ref{lemma_subgraph_to_subset}}\label{proof_lemma_subgraph_to_subset}
 We prove the equality by establishing inequalities in both direction.

\medskip
\noindent {\em Establishing $\min_{S \subset V} \frac{|S|-1}{|E(S)|} \geq \min_{F\subset E} \frac{|V(F)|-c(F)}{|F|}$:} For $S\subset V$ note that $V(E(S)) \subset S$ and $c(E(S)) \geq 1$ and therefore that $\frac{|S|-1}{|E(S)|} \geq \frac{|V(E(S))|-c(E(S))}{|E(S)|}$ with $E(S) \subset E$. Thus, $\min_{S \subset V} \frac{|S|-1}{|E(S)|}$
is minimizing a larger objective function over smaller set compared to $\min_{F\subset E} \frac{|V(F)|-c(F)}{|F|}$. Therefore, the inequality follows immediately. 

\medskip
\noindent {\em Establishing $\min_{S \subset V} \frac{|S|-1}{|E(S)|} \leq \min_{F\subset E} \frac{|V(F)|-c(F)}{|F|}$:} Let $\Fs \subset E$ be a minimizer
of $\min_{F\subset E} \frac{|V(F)|-c(F)}{|F|}$. Let $H = (V(\Fs), \Fs)$. 
By optimality, all connected components of $H$ must be vertex-induced subgraphs of $G$. This is because, if not then it is possible to add edges to $H$ without changing 
the number of vertices or number of connected components in it, which would contradict
optimality. In other words, there exists disjoint subsets $S_i, 1\leq i \leq c(H)$ of $V(\Fs)$ with $V(\Fs) = \cup_{i=1}^{c(H)} S_i$ and $\Fs = \cup_{i=1}^{c(H)} E(S_i)$. 
If $c(H) = 1$, then the inequality follows immediately. If $c(H) \geq 2$, 
denote $H\setminus H_1$ the graph obtained by removing $H_1 = (S_1, E(S_1))$ 
from $H$. Note that $c(H\setminus H_1) = c(H) -1$ and that $c(H_1)=1$. 
By Lemma \ref{lemma_remarkable_identity}, $\forall a,b,c,d \in \Rbb_+^4: \min(\frac{a}{b},\frac{c}{d}) \leq \frac{a+c}{b+d}$. Therefore,
\begin{equation}
    \min\left(\frac{|V(H_1)|-c(H_1)}{|E(H_1)|}, \frac{|V(H\setminus H_1)|-c(H\setminus H_1)}{|E(H\setminus H_1)|}\right) \leq \frac{|V(H)|-c(H)}{|E(H)|}.
\end{equation}
If $H_1$ achieves the minimum on the left hand side, then it concludes the proof. 
If $H\setminus H_1$  achieves the minimum simply iterate the above argument till we
are left with single connected component and that would conclude the proof. 
 
\begin{lemma}\label{lemma_remarkable_identity}
For any $a, b, c, d\in \Rbbp$, 
\begin{align}
 \min\left(\frac{a}{b},\frac{c}{d}\right) & \leq  \frac{a+c}{b+d}.
\end{align}
\end{lemma}
 Let $a,b,c,d \in\Rbbp$. The following
sequence of statements hold leading to the proof of the claim:
\begin{align}
    ad &\leq bc \quad \text{or} \quad bc \leq ad\\
    \min(ad(b+d), cb(b+d))&\leq (a+c)bd\\
    \min\left(\frac{a}{b},\frac{c}{d}\right) &\leq  \frac{a+c}{b+d}.
\end{align}

\section{Proof of Lemma \ref{lem:special.1}}\label{proof_lemma_bound_kappa}
{\em Case of bounded maximum average degree.} Assume $G$ has maximum average degree bounded by $\bar{d}$, where by definition
\begin{equation}
    \bar{d} = \max_{S\subset V}\frac{2|E(S)|}{|S|}.
\end{equation}
Therefore, for any $S \subset V$, $|E(S)| \leq \frac{\bar{d}}{2}|S|$. And there can be 
at most ${|S| \choose 2}$ edges in a graph over vertices $S$, and hence 
$|E(S)| \leq \frac{|S|(|S|-1)}{2}$. Therefore, we obtain
\begin{align}
    \frac{|S|-1}{|E(S)|} &\geq \frac{2}{\bar{d}}\big(1-\frac{1}{|S|}\big) = L_1(|S|),\\
    \frac{|S|-1}{|E(S)|} &\geq \frac{2}{|S|} = L_2(|S|).
\end{align}
Therefore 
\begin{align}
    \frac{|S|-1}{|E(S)|} \geq \min_{x \in \Rbbp}\{\max(L_1(x),L_2(x))\}.
\end{align}
Note that $L_1$ is increasing and bounded whereas $L_2$ is decreasing. Therefore, 
$\max(L_1(x), L_2(x))$ with $x\in \Rbb$ reaches its minimum for $x$ such that $L_1(x) = L_2(x)$ which leads
to minima at $x = \bar{d}+1$. Therefore, we conclude that for all $S \subset V$,
\begin{align}
    \frac{|S|-1}{|E(S)|} & \geq \frac{2}{\bar{d}+1}.
\end{align}

\medskip\noindent
{\em Case of bounded girth.} 
Let $G$ with girth $g > 3$. Note that all subgraphs of $G$ have girth at least $g$. 
The generalised Moore bound (obtained by \cite{alon2002moore}) then gives $\forall S \subset V$:
\begin{align}
    |S| &\geq 1 + d_S\sum_{i=0}^{\frac{g-3}{2}}(d_S-1)^i \quad\quad &\text{if g is odd},\\
    |S| &\geq 2\sum_{i=0}^{\frac{g-2}{2}}(d_S-1)^i, \quad\quad &\text{if g is even}
\end{align}
with $d_S = \frac{2|E(S)|}{|S|}$. We will only keep a weaker version of this bound that does not depend on the parity of $g$. Specifically, for all $S\subset V$:
\begin{align}
    |S| & \geq \big(2\frac{|E(S)|}{|S|}-1\big)^{\frac{g-3}{2}}.
\end{align}
Therefore,  $|E(S)| \leq \frac{1}{2}(|S|^{\frac{2}{g-3}+1}+|S|)$ for all $S \subset V$. Subsequently, 
we have 
\begin{align}
\frac{|S|-1}{|E(S)|} & \geq 2\frac{1-\frac{1}{|S|}}{1+|S|^{\frac{2}{g-3}}} \geq 2\frac{1-\frac{1}{|S|}}{1+N^{\frac{2}{g-3}}}
\end{align}
This bound is clearly increasing with $|S|$. Also note that if $|S| \leq g-1$, the subgraph $(S,E(S))$ can have no cycle and therefore $\frac{|S|-1}{|E(S)|} = 1$. The worse case is therefore attained for $|S| = g$ where we have:
\begin{align}
\frac{|S|-1}{|E(S)|} & \geq \frac{2}{1+N^{\frac{2}{g-3}}}\big(1-\frac{1}{g}\big).
\end{align}

\section{Proof of Lemma \ref{lemma_finite_sample}}\label{proof_finite_sample}
We shall use Hoeffding's inequality: for any bounded random variable $a \leq X \leq b$, 
the deviation of its $n$-empirical average $\overline{X}_n$ computed from in dependant samples is such that for any $t > 0$,
\begin{equation}
    \mathbb{P}(|\Ebb(X)- \overline{X}_n| \geq t) \leq 2 \exp\Big(\frac{-2nt^2}{(b-a)^2}\Big).
\end{equation}
Another version of the equation when $\mathbb{E}(X) > 0$ is as follows, for any $\epsilon > 0$
\begin{equation}
    \mathbb{P}\left(1-\epsilon\leq \frac{\overline{X}_n}{\Ebb(X)}\leq 1+\epsilon\right) 
    \geq 1-2 \exp\Big(\frac{-2n\epsilon^2\mathbb{E}(X)^2}{(b-a)^2}\Big).
\end{equation}
An immediate consequence is that $\hbu^n$ is a good approximation for $\ubf$. For any $e\in E$,
\begin{equation}
    \mathbb{P}\left(1-\epsilon\leq \frac{\hu^n_e}{u_e}\leq 1+ \epsilon \right) \geq 1 - 2\exp{\left(-2n\epsilon^2u_e^2\right)},
\end{equation}
therefore by union bound,
\begin{equation}
    \mathbb{P}\left(\forall e\in E: 1-\epsilon 
    \leq \frac{\hu^n_e}{u_e}\leq 1+\epsilon \right)\geq 1 - 2|E|\exp{\left(-2n\epsilon^2\kappa(G,\bu)\right)}. \label{eq:empirical_distribution}
\end{equation}
Another consequence is that $L_{\hbu^n}$ is a good approximation for $L_{\ubf}$. Indeed, considering the random variable $\Phi(\Pi^\T(\btheta))$ of mean $L_{\bu}(\btheta)$ and of empirical average $L_{\hbu^n}(\btheta) ~=~ \frac1n \sum_{i=1}^n \Phi(\Pi^{\T_i}(\btheta))$ and noting that this variable is bounded as follows $ 0 \leq \Phi(\Pi^\T(\btheta)) \left(\leq \Phi(\btheta)\right) \leq \frac{1}{\kappa_{\bu}} L_{\bu}(\btheta)$, we have 
\begin{align}
    \mathbb{P}\left(1-\epsilon 
    \leq \frac{L_{\hbu^n}(\btheta)}{L_{\bu}(\btheta)}
    \leq 1+\epsilon \right) & 
    \geq 1 - 2 \exp\left(-2 n \epsilon^2\kappa_{\bu}^2 \right).\label{eq:lower}
\end{align}
Regarding $U_{\hbu^n}(\btheta)$, the discussion requires an additional argument because $\frac1n \sum_{i=1}^n \Phi(\Pi^{\T_i}_{\hbu^n}(\btheta))$ is not a sum of independent random variables. Instead, let us focus on the close quantity, $\frac1n \sum_{i=1}^n \Phi(\Pi^{\T_i}_{\bu}(\btheta))$ for which we have $0 \leq \Phi(\Pi^\T_{\bu}(\btheta)) \left(\leq \frac{1}{\kappa_{\bu}} \Phi(\btheta) \right) \leq \frac{1}{\kappa_{\bu}} U_{\bu}(\btheta)$ and therefore,
\begin{align}
    \mathbb{P}\left(1-\epsilon
    \leq \frac{\frac1n \sum_{i=1}^n \Phi(\Pi^{\T_i}_{\bu}(\btheta))}{U_{\bu}(\btheta)}
    \leq 1+\epsilon \right) & 
    \geq 1 - 2 \exp\left(-2 n \epsilon^2 \kappa_{\bu}^2\right).
\end{align}
Fortunately, if \eqref{eq:empirical_distribution} is
satisfied this quantity turns out to be a good approximation of $U_{\hbu^n}(\btheta)$. Indeed, assuming that $ \forall e\in E: 1-\epsilon\leq \frac{\hu^n_e}{u_e} \leq 1+\epsilon$ we have that for all $T\in \Tcal(G)$,
\begin{equation}
   (1-\epsilon)\Pi^T_{\bu}(\btheta) \preceq \Pi^T_{\hbu^n}(\btheta) \preceq  (1+\epsilon) \Pi^T_{\bu}(\btheta)
\end{equation}
therefore by \eqref{eq:monotonicity} and \eqref{submodularity},
\begin{equation}
   (1-\epsilon)\Phi(\Pi^T_{\bu}(\btheta)) \leq \Phi(\Pi^T_{\hbu^n}(\btheta)) \leq  (1+\epsilon) \Phi(\Pi^T_{\bu}(\btheta)),
\end{equation}
which shows,
\begin{equation}
    (1-\epsilon)
    \leq \frac{U_{\hbu^n}(\btheta)}{\frac1n \sum_{i=1}^n \Phi(\Pi^{\T_i}_{\bu}(\btheta))} 
    \leq (1+\epsilon).
\end{equation}
Therefore by union bound,
\begin{equation}
    \mathbb{P}\left((1-\epsilon)^2
    \leq \frac{U_{\hbu^n}(\btheta)}{U_{\bu}(\btheta)} 
    \leq (1+\epsilon)^2
    \right) \geq 1 - (2|E|+2)\exp\left(-2 n \kappa_{\bu}^2 \epsilon^2\right).\label{eq:upper}
\end{equation}
By putting together \eqref{eq:lower} and \eqref{eq:upper}, we obtain
\begin{equation}
   \mathbb{P}\left(
   (1-\epsilon)^{\frac32} 
    \leq \frac{\hPhi_{\hbu^n}(\btheta)}{\Phi_{\bu}(\btheta)} 
    \leq (1+\epsilon)^{\frac32}\right) \geq 1 - (2|E|+4)\exp\left(-2 n \kappa_{\bu}^2 \epsilon^2\right),
\end{equation}
and by arguments of Lemma \ref{lemma_approx}, we can conclude that
\begin{equation}
   \mathbb{P}\left(
   \sqrt{\kappa_{\bu}}(1-\epsilon)^{\frac32} 
    \leq \frac{\hPhi_{\hbu^n}(\btheta)}{\Phi(\btheta)} 
    \leq \frac{(1+\epsilon)^{\frac32}}{\sqrt{\kappa_{\bu}}}\right) \geq 1 - (2|E|+4)\exp\left(-2 n \sqrt{\kappa_{\bu}}^2 \epsilon^2\right),
\end{equation}
This completes the proof of Lemma \ref{lemma_finite_sample}.

\section{Proof of Lemma \ref{lemma_eff_res}}\label{proof_lemma_eff_res}

\medskip
\noindent{\em Bounded degree graph $G$.} First assume that $G$ has maximum degree $d$. 
Consider any edge $e =(s,t) \in E$. Denote $\Ncal(s), \Ncal(t) \subset V$ the neighbours of $s$ and $t$. Consider current $\iota: V \times V \to \Rbb$ which is a solution of optimization
problem corresponding to effective resistance as defined in \eqref{eq:eff.res}. 
By definition, we have that the effective resistance $u_e$ for $e \in E$ is
given by
\begin{align}\label{proof_ineq_effective}
    u_e & = \sum_{(u,v) \in E} \iota(u,v)^2 \nonumber \\
        & \geq \iota(s,t)^2 + \sum_{u \in \Ncal(s)\setminus\{t\}} \iota(s,u)^2 
            + \sum_{u \in \Ncal(t)\setminus\{s\}} \iota(u,t)^2.
\end{align}
By constraints of the optimization problem, the sum of currents entering source $s$ and leaving sink $t$ is equal to 1 (whereas it is null for isolated vertices). 
Therefore, focusing on $s$, we have 
$\sum_{u \in \Ncal(s)\setminus\{t\}}|\iota(s,u)| \geq  1 - |\iota(s,t)|$. 
By applying Cauchy Schwarz inequality, we have that
\begin{equation}
\big(\sum_{u \in \Ncal(s)\setminus\{t\}}\iota(s,u)^2\big) \times 
\big(\sum_{u \in \Ncal(s)\setminus\{t\}}1^2\big) \geq (1-|\iota(s,t)|)^2.
\end{equation}
Recall that $G$ has maximum vertex degree $d$ and 
therefore $|\Ncal(s)\setminus \{t\}| \leq d-1$. Therefore, 
\begin{equation}
    \sum_{u \in \Ncal(s)\setminus\{t\}} \iota(s,u)^2 \geq \frac{(1-|\iota(s,t)|)^2}{d-1}.
\end{equation}
Because the same holds for the term 
$\sum_{u \in \Ncal(t)\setminus(s)}\iota(u,t)^2$, 
we obtain from (\ref{proof_ineq_effective}) that
\begin{equation}
    u_e \geq \iota(s,t)^2 +(1-|\iota(s,t)|)^2 \frac{2}{d-1}.
\end{equation}
This expression holds for all possible values of $\iota(s,t)$. 
We note that for any given $\lambda \in \Rbbp$, 
\begin{align}\label{eq:eff.ref.misc}
    \inf_{x \in \Rbb} x^2 + (1-x)^2 \lambda & \geq \frac{\lambda}{1+\lambda}.
\end{align}
Therefore, we conclude that for graph $G$ with bounded degree $d$,
\begin{equation}
    u_e \geq \frac{2}{d+1}.
\end{equation}

\medskip
\noindent{\em Graph $G$ with girth $g$.} We now assume that $G$ has girth $g$. As
before, let $e=(s,t) \in E$. Denote $G\setminus\{e\} = (V,E\setminus\{e\})$ 
the graph obtained by removing edge $e$ from $G$. For $0\leq k\leq g-2$, 
we define 
\begin{align}
E_k & = \{(u,v)\in E : d_{G\setminus\{e\}}(s,u) = k, d_{G\setminus\{e\}}(s,v) = k+1\}, 
\end{align}
where $d_{G\setminus\{e\}}(s,u)$ denotes the shortest path distance between vertices $s, u$
in graph $G$ excluding edge $e$. That is, $E_k$ is the set of edges connecting vertices at distance $k$ from $s$ in $G\setminus\{e\}$ to vertices at distance $k+1$ from $s$ in $G\setminus\{e\}$. Since $k \leq g-2$, all $E_k$ are disjoint and hence
current $\iota$ satisfies
\begin{align}\label{eq girth basic}
    u_e & \geq \iota(s,t)^2 + \sum_{k=0}^{g-2} \sum_{(u,v)\in E_k} \iota(u,v)^2.
\end{align}
For $0\leq k\leq g-2$, note that $E_k \cup \{e\}$ defines a cut of $G$. 
Therefore by Kirchoff's law $\sum_{(u,v) \in E_k}|\iota(u,v)| \geq 1-|\iota(s,t)| $. 
Using Cauchy-Schwartz inequality, we obtain:
\begin{align}
    \big(\sum_{(u,v)\in E_k} \iota(u,v)^2\big) \times \big(\sum_{(u,v)\in E_k}1^2\big) \geq (1-\iota(s,t))^2.
\end{align}
By summing-up all inequalities, we obtain
\begin{align}
   \big(\sum_{k=0}^{g-2}\sum_{(u,v) \in E_k} \iota(u,v)^2) & \geq (1-|\iota(s,t)|)^2 \big(\sum_{k=0}^{g-2}\frac{1}{|E_{k}|}\big).
\end{align}
Note that if a sequence $(m_k) \geq 0$ respects $\sum_{k =1}^{l} m_k \leq |E|$ then, 
$\sum_{k =1}^{l} \frac{1}{m_k} \geq \frac{l^2}{|E|}$. Therefore, because all $E_{k}$ are disjoint, $\sum_{k=0}^{g-2}\frac{1}{|E_{k}|}\geq \frac{(g-1)^2}{|E|}$. Inserting
this in \eqref{eq girth basic}, we obtain
\begin{align}
u_e & \geq \iota(s,t)^2 + (1-\iota(s,t))^2 \frac{(g-1)^2}{|E|}.  
\end{align}
Using \eqref{eq:eff.ref.misc}, we obtain 
\begin{align}
    u_e & \geq \frac{1}{1+\frac{|E|}{(g-1)^2}}.
\end{align}
This completes the proof of Lemma \ref{lemma_eff_res}.

\section{Proof of Theorem \ref{theorem_partitionning}}\label{proof_partitionning}

The proof follows by establishing that by definition if $\brho \in \Pcal(\Prt_k(G))$ is an $(\epsilon,k)$-partitioning of $G$, it implied that
\begin{align}
    \kappa(G,\brho) & \geq 1-\epsilon, 
\end{align}
Indeed, by definition of $(\epsilon, k)$ partition, 
we have that for any $e \in E$, 
\begin{align}
    \rho_e & = \Ebb_{\Hs \sim \brho}[\one(e \in \Hs)] \geq 1-\epsilon.
\end{align}
Subsequently, using arguments identical to that for proof of Lemma \ref{lemma_approx}, it follows
that $\hPhi_{\brho}(\btheta)$ is $1/\sqrt{\kappa(G,\brho)}$ approximation. That is, 
    \begin{equation}
    \sqrt{1-\epsilon}\leq \frac{\Phi(\btheta)}{\hPhi_{\brho}(\btheta)}\leq \frac{1}{\sqrt{1-\epsilon}}.
\end{equation}
This completes the proof of Theorem \ref{theorem_partitionning}.

\end{document}